\documentclass[a4paper,12pt,onecolumn,oneside,iopams,setstack,amsmath,amssymb,amsfonts]{iopart}

\usepackage[pdftex]{hyperref}                    
\usepackage[pdftex]{graphicx}
\usepackage{cite}
\usepackage{dcolumn}                             
\usepackage{bm}                                  
\usepackage[verbose,tmargin=3cm,bmargin=3cm,lmargin=3cm,rmargin=3cm]{geometry}
\usepackage[switch,running]{lineno}              

\newcommand{\dif}{\mathrm{d}}                    
\newcommand{\me}{\mathrm{e}}                     
\newcommand{\mpi}{\mathrm{\pi}}                  

\begin{document}


\title[A computational physics project for radiation trapping in 1D]
      {\Large
       Radiation trapping in 1D using the Markov chain formalism: \\
       A computational physics project}

\author{A~R~Alves-Pereira$^1$, E~J~Nunes-Pereira$^{1, \, \dagger}$,
  J~M~G~Martinho$^2$ and M~N~Berberan-Santos$^2$}

\address{$^1$ Universidade do Minho, Escola de Ci\^{e}ncias,
  Centro de F\'{i}sica, 4710-057 Braga, Portugal}

\address{$^2$ Centro de Qu\'{i}mica-F\'{i}sica Molecular, Instituto
  Superior T\'{e}cnico, 1049-001 Lisboa, Portugal}

\ead{$^{\dagger}$~epereira@fisica.uminho.pt}

\date{\today}

\begin{abstract}

A computational model study for complete frequency redistribution linear incoherent two-level
atomic radiation trapping in optically dense media using the multiple scattering representation is
presented. This model study discuss at length the influence of the spectral distributions, overall
opacity and emission quantum yield to trapping distorted ensemble quantities stressing physical
insight and with a non-specialist audience in mind. Macroscopic reemission yield, lifetime, steady
state spectra and spatial distributions are calculated as a function of intrinsic emission yield,
opacity and external excitation mode for Doppler, Lorentz and Voigt lineshapes. The work could
constitute the basis for a final undergraduate or beginning graduate project in computational
physics instruction and implements the analytical developments of the previous instalment of this
contribution.

\end{abstract}

\maketitle

\section{Introduction}\label{Introduction}

In optically thick media, electronic excitation energy can undergo several reabsorption and
reemission events before either escaping to the exterior or being converted in thermal energy by
means of collisional deactivation. This \textit{resonant radiation trapping} is important in areas
as diverse as stellar atmospheres~\cite{Mihalas1978:book}, plasmas and atomic vapours
luminescence~\cite{Molisch1998:book}, terrestrial atmosphere and ocean
optics~\cite{Thomas1999:book}, molecular luminescence~\cite{Berberan-Santos1999:chapter}, infrared
radiative transfer~\cite{Modest2003:book} and cold atoms~\cite{Bardou2002:book}.

The starting point of the majority of the incoherent radiation trapping models is
the~\textit{Holstein-Biberman equation} which is a Boltzmann-type integro-differential equation
describing the spatial and temporal evolution of the excited state number density. The previous
instalment~\cite{PartI} of this contribution outlined the two alternative ansatze commonly used to
obtain solutions for the classical trapping problem, Holstein's original exponential mode expansion
and the so called Multiple Scattering Representation~(MSR). The~MSR solution was given a simple
stochastic formulation and trapping dependent quantities~(overall relaxation parameters such as
ensemble emission yield and lifetime, time-resolved and steady-state spatial distributions as well
as spectra) were calculated with the Holstein fundamental mode singled out. This instalment will
now use a simple Markov chain algorithm to quantify incoherent trapping in a computational model
study for two-level atomic models. It could easily be adapted into a computational physics project
valuable in the context of atomic or computational physics instruction for final undergraduate and
beginning graduate students. With this objective in mind, trapping dependent quantities are
estimated in a unidimensional geometry for a single line with Doppler, Lorentz and Voigt spectral
distributions. The excitation relaxation dynamics is considered for conditions mimicking electron
impact as well as photoexcitation. A thorough discussion of the conditions for which the use of
Holstein's fundamental mode alone is a tolerable approximation is included.

\Sref{Dynamics of incoherent trapping} discusses the way the dynamics of incoherent trapping is
taken into account within the framework of the~MSR. It gives explicit expressions for the overall
relaxation parameters as well as steady-state spectra and spatial distribution summarizing the
previous instalment of this work. \Sref{Markov stochastic algorithm} explains the rationale as well
as critical implementation details of the Markov stochastic simulation algorithm, the one chosen in
this model study. \Sref{Results and discussion} presents the results and their discussion at
length. Particular emphasis is placed in the discussion of the physical implications of the
spectral distribution as well as quantifying the relative contribution of the fundamental mode.
\Sref{Conclusions} summarizes the main conclusions. Finally, the Appendices show implementation
fine details and add some possible routes for complementing the work presented here.

\section{Dynamics of incoherent trapping}\label{Dynamics of incoherent trapping}

Radiation trapping studies should be cast in dimensionless coordinates since this increases
computational efficiency and, more important, defines characteristic scales or universal
conditions. The quantities most directly amenable to define characteristic scales in trapping are
time, distance and optical frequency. The scaled time is~$t=\Gamma t^{\prime }$, where~$\Gamma$ is
the global deactivation rate constant. The dimensionless distance is sometimes called the
\textit{opacity} or \textit{optical density} and can be defined, along a given pathlength~$l$ and
for homogeneously distributed species, as~$r=k\left( x\right) / \Phi \left( 0\right)=k_{0} \, \Phi
\left( x\right) / \Phi \left( 0\right)$. $x$~is used to represent the optical frequency~(see
below).~$k\left( x\right) $ is the single line monochromatic opacity~($k_{0}$ is the corresponding
center-of-line value) and the absorption lineshape is given by the normalized spectral
distribution~$\Phi \left( x\right)$~(so that~$\int_{-\infty }^{+\infty }\Phi \left( x\right) \,\dif
x = 1$). For two-level atomic models the intrinsic, trapping undistorted, spectra can be written as
a function of a dimensionless optical frequency defined as~$x=\frac{\nu-\nu_0}{\Delta\nu^{D}}$.
This is a normalized difference to the center of line frequency, where~$\Delta\nu^{D}$ stands for
the~FWHM of the Doppler distribution at each given temperature. For two-level atomic models, it is
common to describe the absorption lineshape by Doppler~\textendash ~$\Phi _{D}\left( x\right) =
\frac{1}{\sqrt{\mpi}}\me ^{-x^{2}}$~\textendash , Lorentz~\textendash ~$\Phi_{L}\left( x\right)
=\frac{1}{\mpi}\frac{1}{1+x^{2}}$~\textendash , or Voigt~\textendash~$\Phi _{V}\left(x\right)
=\frac{a}{\mpi ^{3/2}}\int_{-\infty }^{+\infty }\frac{\me ^{-u^{2}}} {a^{2}+\left( x-u\right)
^{2}}\,\dif u$~\textendash ~spectral distributions. The Doppler distribution allow us to single out
the pure Doppler broadening from the other broadening mechanisms while the Lorentz and Voigt's
distributions are the ones to be used in pure radiation damping and combined radiation and
collision broadening conditions, respectively. In this last case, $a=\sqrt{\ln \left( 2\right)
}\frac{\Delta \nu^{L}}{\Delta \nu ^{D}}$ is the Voigt characteristic width, the relative Lorentz
over Doppler spectral width and implicitly dependent upon both temperature and vapour pressure via
the dependence on collisional cross-section values.

In the~MSR ansatz for linear incoherent trapping with negligible time-of-flight for in-transit
radiation, the spatial and temporal relaxation dynamics for excitation is given by

\begin{equation}
\label{MSRSol}
n\left( \bi{r},t\right) =\sum_{n} a_{n} p_{n}\left( \bi{r}\right) g_{n}\left(
t\right)
  \mbox{,}
\end{equation}

where $n$ stands for the \textit{generation number} of excited species~(primordial excitation
creates first generation, the trapping of this generation's reemission creates second generation
and so forth; one can envisage each generation as the result of~$n-1$ previous scattering events of
resonant radiation), $a_{n}$ is the population efficiency for each generation, and $p_{n}\left(
\bi{r}\right)$ and~$g_{n}\left( t\right)$ are the (normalized) spatial and temporal excited species
distributions.

From a practical point of view, the most important quantities are the macroscopic ensemble
relaxation parameters, overall reemission yield~$\phi$ and mean scaled lifetime~$\tau$, and the
steady state emission spectrum~$I^{SS}\left( x\right)$. These were derived in a previous
contribution and are given by~\cite{PartI}:

\begin{equation}
\label{phi}
\phi = \sum_{n=1}^{m-1}q_{n}a_{n}+\frac{q_{nc}a_{nc}}{1-\alpha
   _{nc}}
  \mbox{,}
\end{equation}

\begin{equation}
\label{Tau}
\tau =\frac{\sum_{n=1}^{m-1}nq_{n}a_{n}+\frac{q_{nc}a_{nc}}{\left( 1-\alpha
_{nc}\right) ^{2}}\left[ m\left( 1-\alpha _{nc}\right) +\alpha _{nc}\right] }{\phi }
  \mbox{,}
\end{equation}

and

\begin{equation}
\label{ISS}
I^{SS}\left( x\right) =\frac{\sum_{n=1}^{m-1}q_{n}^{\Omega }\left( x\right)
a_{n}+\frac{q_{nc}^{\Omega }\left( x\right) a_{nc}}{1-\alpha _{nc}}}{\int_{-\infty }^{+\infty
}\left[ \sum_{n=1}^{m-1}q_{n}^{\Omega }\left( x\right) a_{n}+\frac{q_{nc}^{\Omega }\left( x\right)
a_{nc}}{1-\alpha _{nc}}\right] \Phi \left( x\right) dx}\Phi \left( x\right)
  \mbox{.}
\end{equation}

In these,~$m$ stands for the first generation number that can be considered
nonchanging~(subscript~$nc$ is a remainder for \textit{nonchanging}; see~\cite{PartI}) and
$\alpha_{n}\equiv \frac{a_{n+1}}{a_{n}}$ is the mean trapping or reabsorption probability. Finally,

\begin{equation}
\label{EscapeOmega}
q_{n}^{\Omega }\left( x\right) =\int_{\Omega }\! \int_{V}\! \me ^{-\Phi \left(
x\right) r}p_{n}\left( \bi{r}\right) \dif \bi{r}\dif S
  \mbox{,}
\end{equation}

is the mean escape probability in the direction defined by the solid angle~$\Omega$ which, for a
left escape from a unidimensional geometry~(see next section), is just

\begin{equation}
\label{EscapeOmega1D}
q_{n}^{\Omega }\left( x\right) = \int \! \me ^{-\Phi \left( x\right)
r}p_{n}\left(r\right) \dif r
  \mbox{,}
\end{equation}

if one decides to start the opacity scale on the left side of the cell.

It is also informative to know the steady-state spatial distribution, which can be cast
as~\cite{PartI}:

\begin{equation}
\label{DistributionSS} n^{SS}\left( \bi{r}\right) = \frac{\sum_{n=1}^{m-1}a_{n}p_{n}\left( r\right)
+\frac{a_{nc}}{1-\alpha _{nc}}p_{nc}\left( r\right) }{\sum_{n=1}^{m-1}a_{n}+\frac{a_{nc}}{1-\alpha
_{nc}}}
  \mbox{.}
\end{equation}

In the above expressions it is important to recognize that the trapping dynamics can be factored
out into a generation varying part and another corresponding to generations that have the same
spatial distribution~(the fundamental mode) and the nonchanging part can be expressed explicitly as
an analytical sum. The dynamics for this nonchanging part corresponds only to an attenuation of
overall excitation in going from one generation to the next by a fixed~$\alpha_{nc}$ factor and, as
a result, the contribution of this part corresponds to a monoexponential relaxation with a trapping
dependent effective decay constant.

\section{Markov stochastic algorithm}\label{Markov stochastic algorithm}

The stochastic formulation of the Multiple Scattering Representation~(MSR) allows to quantify the
trapping influence on observables essentially trough the estimation of mean reabsorption and escape
probabilities. In this respect, it is simpler and more amenable to discussion at an elementary
level than the alternative Holstein multiexponential expansion. The~MSR numerical implementation is
straightforward and can be easily discussed at a level that stresses physical intuition without
actually getting bogged down in all the technicalities of the troublesome estimation of Holstein
modes other than the fundamental. We have furthermore decided to use an unidimensional geometry,
driven by the motivation to do a simple mimic of a cylindrical tube of a discharge fluorescence
lamp and by the desire to keep computational detail and power adequate to the proposed audience of
final undergraduate and beginning graduate students. In this~1D case the opacity scale comes
naturally as the opacity along the cylindrical axis.

\begin{figure}
  \begin{center}
  \includegraphics[width=12.5cm,keepaspectratio=true]{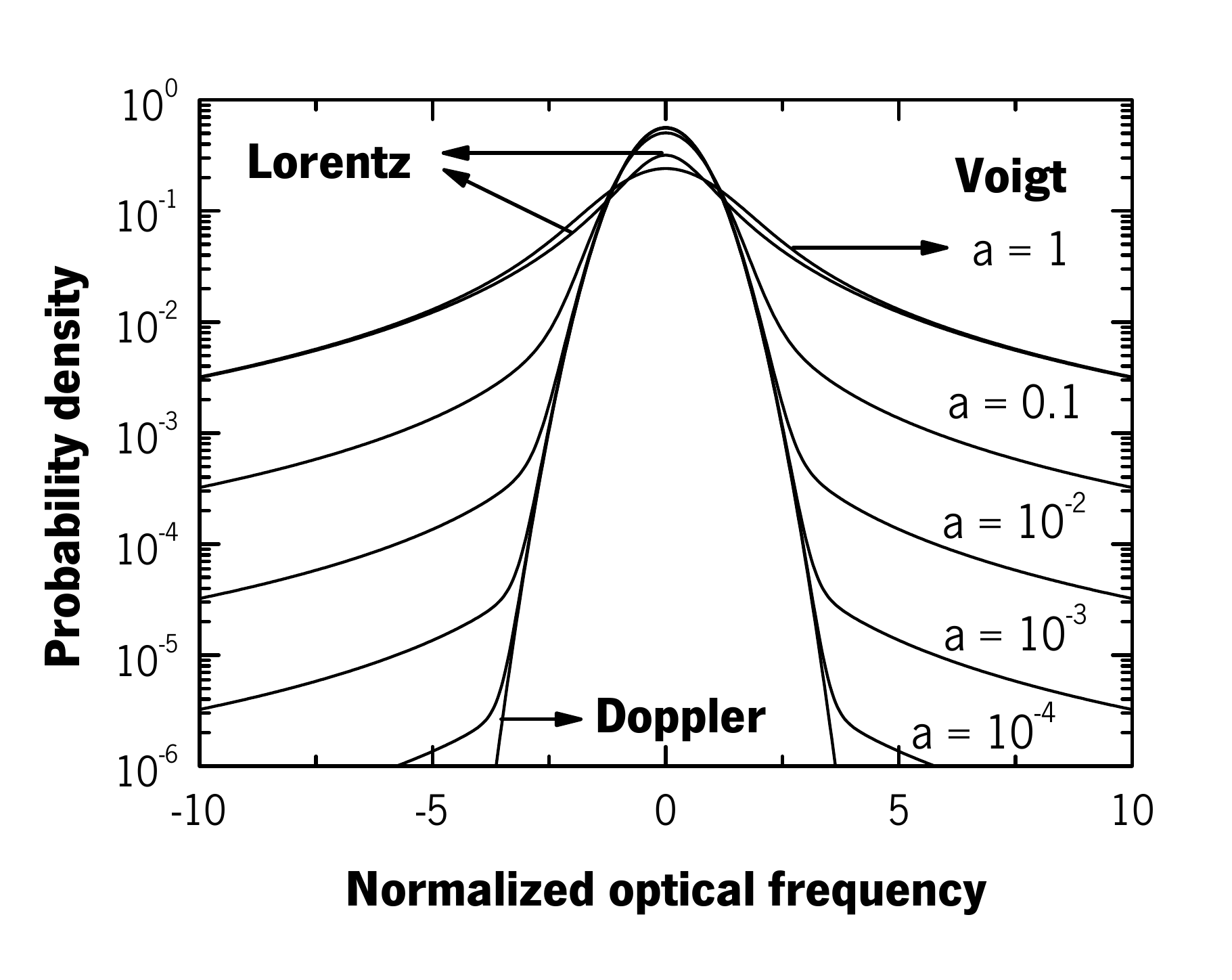}
  \end{center}
  \caption{\label{Fig-Spectra}
    Doppler, Lorentz and Voigt lineshapes.}
\end{figure}

The model used is a two-level single line atomic model with Doppler, Lorentz or Voigt spectral
distributions and particular attention is paid to the influence of the lineshape on the trapping
efficiency. The Voigt case will illustrate the fact that a continuous variation of the
characteristic width parameter will map the Doppler into the Lorentz distributions by changing the
relative importance of the Lorentz-like wings over the Doppler-like core of the
distribution~(\fref{Fig-Spectra}). Accordingly, the ability to compute the Voigt line to machine
precision is mandatory and~\ref{Numerical Voigt distribution} gives some implementation details in
this respect.

Complete frequency redistribution conditions are used~\cite{Molisch1998:book}, which means that the
number of collisions during the lifetime of excited atoms is sufficiently high to render the
reemitted photon's frequency completely uncorrelated with the frequency of the previously absorbed
photon. In these conditions, the absorption and emission lineshapes coincide and the jump length
distribution of the excitation random trajectory is independent of past history. This makes the
formalism of Markov processes~\cite{vanKampen1992:book&Gardiner1985:book} especially adequate. Its
rationale for the~MSR implementation can be cast in the following way. The~1D cell is divided into
several bins, each corresponding to a \textit{pure state} of the system and characterized by a mean
probability that the excitation resides in that state. The system dynamics corresponds to the
evolution of the probability of excitation being inside each state. The stochastic process is
completely specified by~(i)~a column vector with the (normalized) spatial probability distribution
of the first generation species, $\bi{p}_{1}=\left[ p_{1}^{i}\right] $, and~(ii)~a transition
matrix $\bi{P}=\left[ p^{ij}\right] $, whose entries are the \textit{one-step} transition
probabilities between states~$i$ and~$j$. For complete frequency redistribution, there is an
absence of memory effects~(homogeneous chain) meaning that the transition probability between
individual states depends only upon their relative opacity distance~(it is independent of
generation number and thus computed only once). The spatial distribution functions for all the
generations are calculated from the previous generation by:

\begin{equation}
\label{MK-p} \bi{p}_{n+1} = \bi{P} \bi{p}_{n}
  \mbox{.}
\end{equation}

The sample cell is divided into $h$-length bins and the transition matrix elements are therefore
given for a 1D geometry by~\cite{Nunes-Pereira2004:PhysRevLett.93.120201}:

\begin{equation}
\label{Transition-pij} p^{ij} \simeq \frac{1}{2}h\int_{-\infty }^{+\infty } \Phi^{2}\left( x\right)
\me^{-\Phi \left( x\right) \left\vert \bi{r}_{i}-\bi{r}_{j}\right\vert } \dif x
  \mbox{,}
\end{equation}

corresponding to the Beer-Lambert law weighted according with the emission lineshape for an
individual reemission-reabsorption~(scattering) event. The integration takes into account all the
possible emission frequencies, $1/2$ is the left or right emission direction probability for a 1D
geometry and it was assumed that the bin width is controlled in order to attain a satisfactory
precision.

The complete specification of the Markov process is achieved once one specifies the initial spatial
distribution~$\bi{p}_{1}$. Two different cases were considered, one for homogeneous initial
excitation~(trivial) and another mimicking photoexcitation with the reabsorption undistorted line.
For photoexcitation from the \textit{left} side of the 1D cell,

\begin{equation}
  \label{Initial-p1i}
  p_{1}^{i} \simeq h\int_{-\infty }^{+\infty } \Phi^{2}\left( x\right)
  \me^{-\Phi \left( x\right) {r}_{i} } \dif x
  \mbox{,}
\end{equation}

which afterwards must be properly normalized.

The binning of the spatial excitation distributions corresponds to the substitution of a continuous
distribution for its discretized version, effectively transforming a continuous process into a
discrete one realized in a lattice. It is therefore the Markov equivalent of a random walk defined
over a regular spaced lattice~\cite{Hughes-Vol1-1995:book}. The bin width~(or the number of cells)
is the critical parameter for the Markov algorithm. We have conducted several tests and found
advisable to have a maximum bin size of $0.05$ in an opacity scale. Otherwise, numerical artifacts
associated with substituting the actual excitation migration for the jump between the mean
coordinates of each bin of sample cell could exist. These were found to be the more important the
higher the overall opacity.

We have mentioned in the beginning of this section that the Markov implementation of the~MSR model
aims at estimating mean reabsorption and escape probabilities. The reabsorption probabilities are
estimated with the following procedure. Each time~\eref{MK-p} is used, the fraction of the
excitation remaining inside sample cell gives the $n^{th}$ mean reabsorption probability $\alpha
_{n}^{T}$. The excitation column vector is then (re)normalized and the process repeated. From the
values of this parameter for each generation, the trapping population efficiency is
$a_{n}^{T}=\prod\limits_{n=1}^{n-1}\alpha _{n}^{T}$. This procedure is the equivalent of an
importance sampling method~(see~\ref{Monte Carlo simulation}), in which it is assumed a unit
intrinsic reemission yield in~\eref{Transition-pij}. The influence of the actual value of this
yield~($\phi _{0}=\frac{\Gamma _{r}}{\Gamma _{r}+\Gamma _{nr}}$, the ratio of radiative over
radiative plus nonradiative relaxation rate constants) is introduced analytically. Using the
notation developed in the previous instalment, the Markov algorithm directly estimates~$a_{n}^{T}$,
each generation population efficiency due to trapping (and geometry) alone, and the actual
population efficiencies were then given by~$a_{n}=a_{n}^{T}\phi _{0}^{n-1}$~\cite{PartI}. As for
the escape probabilities, we have chosen the following implementation of~\eref{EscapeOmega1D}. The
monochromatic left escape probability is obtained from

\begin{equation}
\label{MK-q} q_{n}^{\Omega }\left( x\right) = \bi{Q}\left( x\right) \bi{p}_{n}
  \mbox{,}
\end{equation}

where $\bi{p}_{n}$ is the spatial distribution and $\bi{Q}\left( x\right)$ is the escape matrix,
whose entries are finally

\begin{equation}
  \label{Transition-q}
  q^{i\Omega }\left( x\right) \simeq \frac{1}{2}
  \me^{-\Phi \left( x\right) r_{i} }
  \mbox{,}
\end{equation}

which comes directly from Beer-Lambert law.

\Eref{MK-p} is equivalent to linear response theory and therefore the evolution of the spatial
excitation is given as a convolution integral between the \textit{excitation profile} $\bi{p}_{n}$
with the \textit{delta response function} given by~\eref{Transition-pij}. This renders the Markov
approach more efficient since this convolution can be easily made using FFT
algorithms~\cite{FFTW3:2005} paying attention to zero pad to double size the column vector
containing the excitation distribution in order to avoid wrap-around effects due to the cyclic
convolution~\cite{NR-F77-2nd:book&NR-F90-2nd:book}. The speed-up factors could rise up to several
orders of magnitude~(roughly~$50$ to~$400$ times for a number of Markov states of~$2\,000$
to~$200\,000$) making the FFT convolution the recommended implementation of the Markov algorithm.

\section{Results and discussion}\label{Results and discussion}
\subsection{Ensemble relaxation}\label{Ensemble relaxation}

\begin{figure}
  \begin{center}
  \includegraphics[width=12.5cm,keepaspectratio=true]{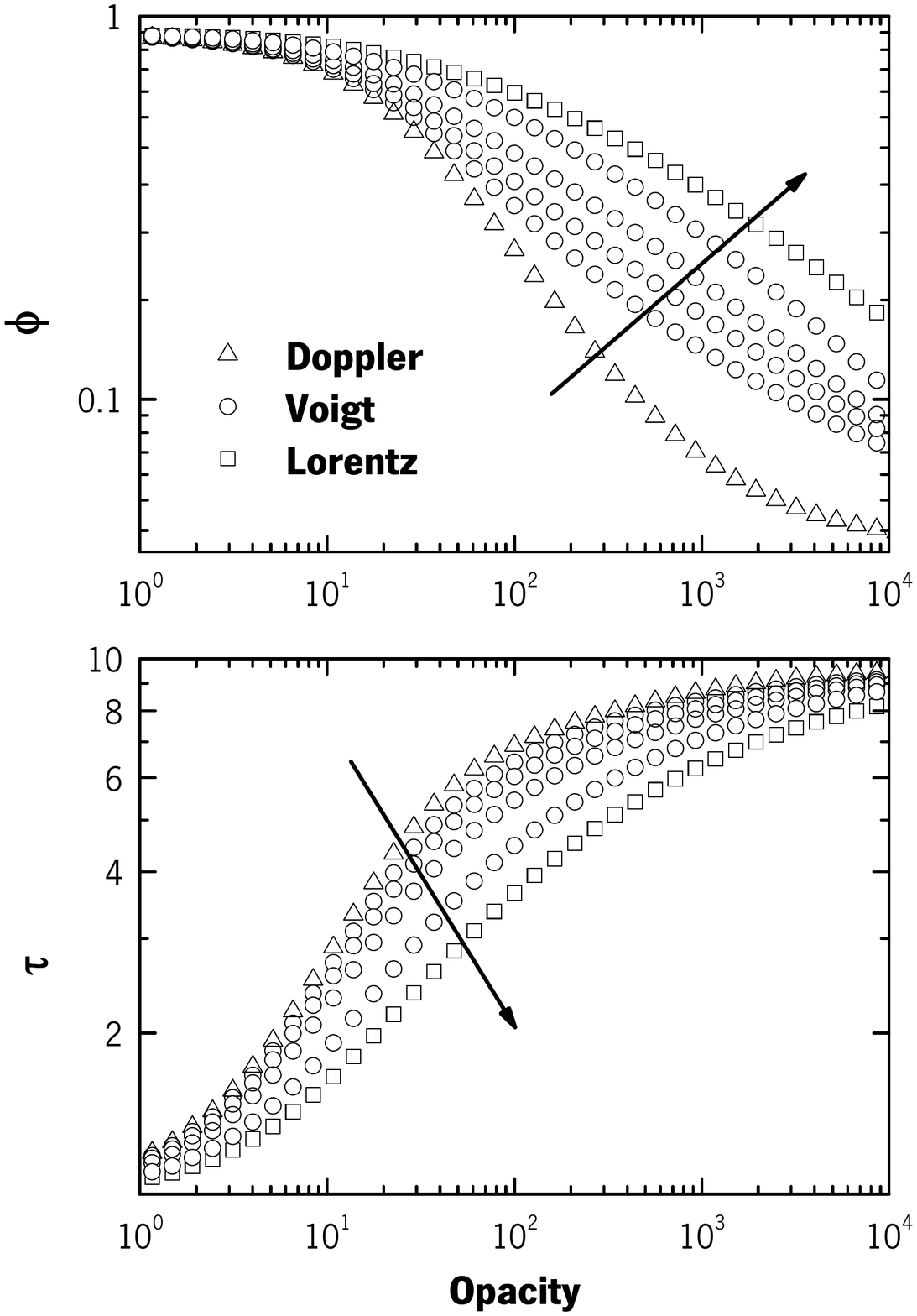}
  \end{center}
  \caption{\label{Fig-All}
    Reemission yield~$\phi$ and mean scaled lifetime~$\tau$, with $\phi_0=0.90$ for Doppler,
    Voigt $a=0.05$, $0.1$, $0.2$ and $0.5$ and Lorentz lineshapes~(direction shown by arrow).}
\end{figure}

\Fref{Fig-All} shows the overall relaxation parameters for an initial homogeneous excitation and an
intrinsic quantum yield of~$\phi_{0}=0.9$ as a function of opacity for Doppler, Voigt and Lorentz
lineshapes. The higher the opacity the more important trapping is, with the following
implications:~(i)~an increase of the mean relaxation time~(equivalent to a mean number of
scattering events before escape) and (ii)~a decrease of the ensemble reemission yield~(the fraction
of original excitation that eventually comes out; an increased importance of trapping translates
into additional possibilities of nonradiative relaxation). The Voigt continuous transition from
Doppler into Lorentz is evident as well as the relative importance of core and wings of the
distributions. The higher the Lorentz character of the spectra the higher the weight of the wings
and the higher the escape probabilities~(lower mean lifetime and higher macroscopic emission
yield).

\begin{figure}
  \begin{center}
  \includegraphics[width=12.5cm,keepaspectratio=true]{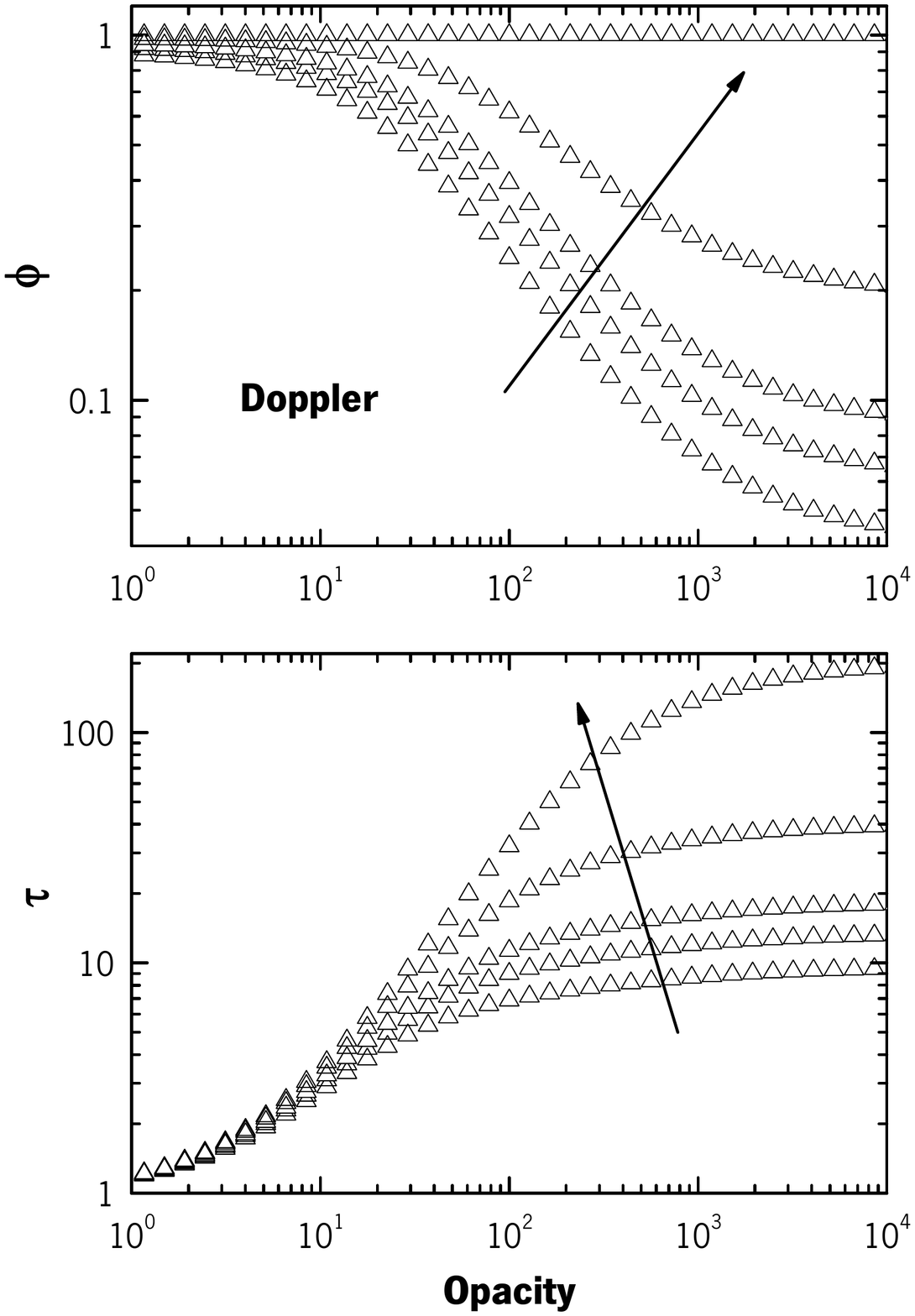}
  \end{center}
  \caption{\label{Fig-Dop}
    Reemission yield~$\phi$ and mean scaled lifetime~$\tau$ for the Doppler lineshape,
    with $\phi_0$ values of $0.90$, $0.93$, $0.95$, $0.98$, and $1.0$~(direction shown by arrow).}
\end{figure}

\begin{figure}
  \begin{center}
  \includegraphics[width=12.5cm,keepaspectratio=true]{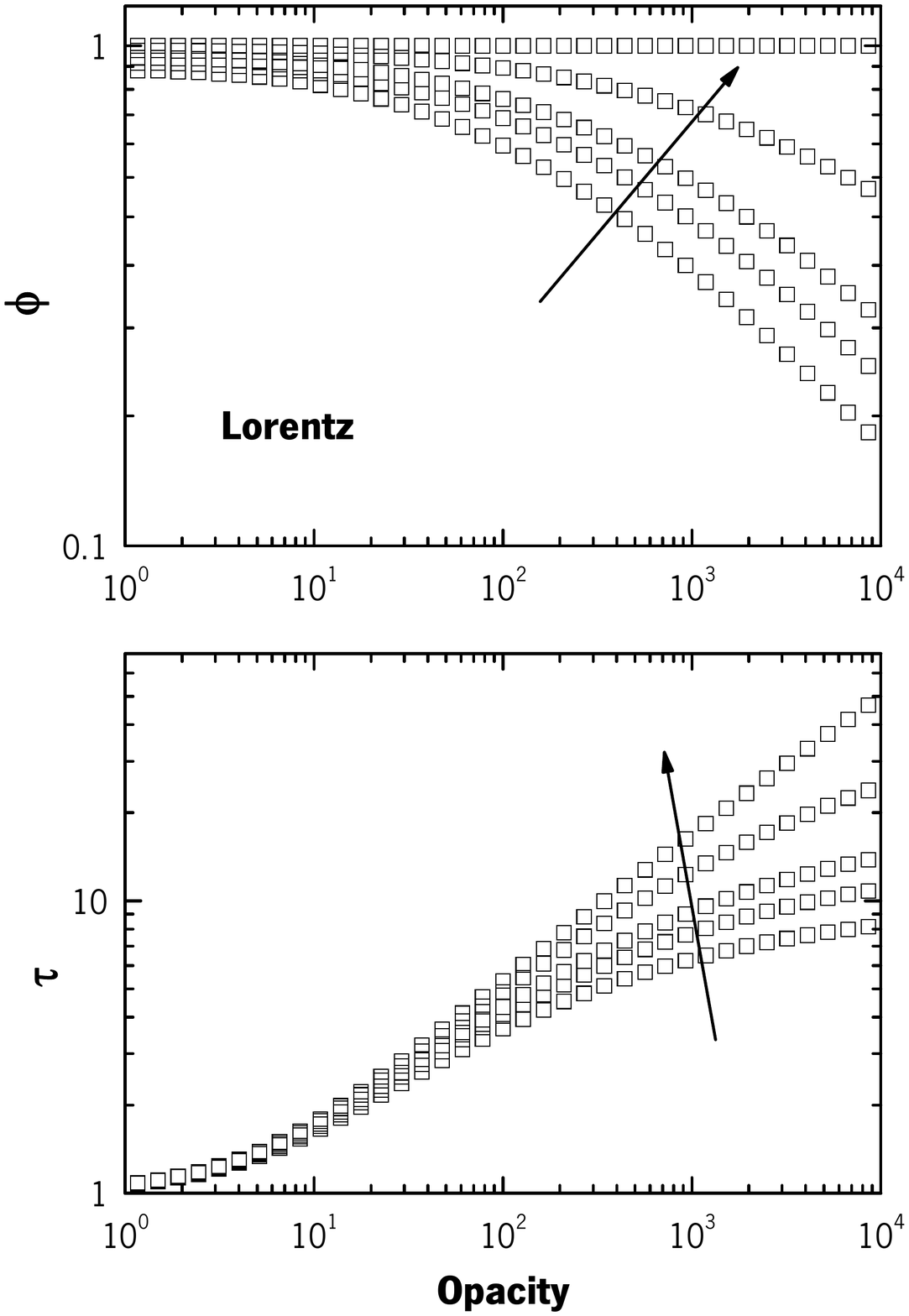}
  \end{center}
  \caption{\label{Fig-Lor}
    Reemission yield~$\phi$ and mean scaled lifetime~$\tau$ for the Lorentz lineshape,
    with $\phi_0$ values of $0.90$, $0.93$, $0.95$, $0.98$, and $1.0$~(direction shown by arrow).}
\end{figure}

Figures~\ref{Fig-Dop} and \ref{Fig-Lor} show the mean scaled lifetime and reemission yield for
homogeneous initial excitation for the Doppler and Lorentz limiting spectral distributions and for
several values of the intrinsic quantum yield. Three main conclusions can be drawn from these
results. First of all, the two most important parameters controlling trapping efficiency are the
spectral distribution and the value of~$\phi_{0}$. The higher the overall opacity the more
difficult is the escape of radiation for Doppler-like distributions and the more important the
escape from the Lorentz-like wings of the distribution. Second, trapping for Doppler-like
distributions is much more efficient since, especially in high opacity cases, the escape of
excitation at optical frequencies far enough from the line center frequency is reduced due to the
extremely small probability of reemission at those frequencies~(for unit reemission yield, the
lifetime for the higher opacity is about~$200$ for Doppler and only about~$15$ for Lorentz).
Finally, under conditions rendering trapping efficient, the~$\phi_{0}$ value is of paramount
importance; for unit intrinsic reemission probability all the excitation will eventually come
out~(thus giving a simple check for consistency of computation) but, as soon as~$\phi_{0}$ is
smaller than one, each new scattering event gives the excitation another chance of thermal
degradation~(in~\eref{phi} each generation contribution is intrinsically dependent upon~$\phi
_{0}^{n-1}$). Note in these figures that there is a strong dependence of the relaxation parameters
with the~$\phi_{0}$ value, especially for the more trapping influenced Doppler case.

All these conclusions are important in the discussion of atomic vapours ensembles for lighting
applications, either electric discharge lamps or plasma display panel~(PDP) devices. Better
performance is achieved with higher macroscopic reemission yields. On top of that, the increase of
the overall opacity is in principle desirable since this is related with the increase of the number
of excited species. In a crude first order approximation, one can assume the overall lamp
efficiency to be directly proportional to the product of~$\phi$ times the overall opacity~(directly
proportional to initial excitation density):

\begin{equation}
\label{phi_lamp}
   \Psi \propto \phi \times r
  \mbox{.}
\end{equation}

\begin{figure}
  \begin{center}
  \includegraphics[width=12.5cm,keepaspectratio=true]{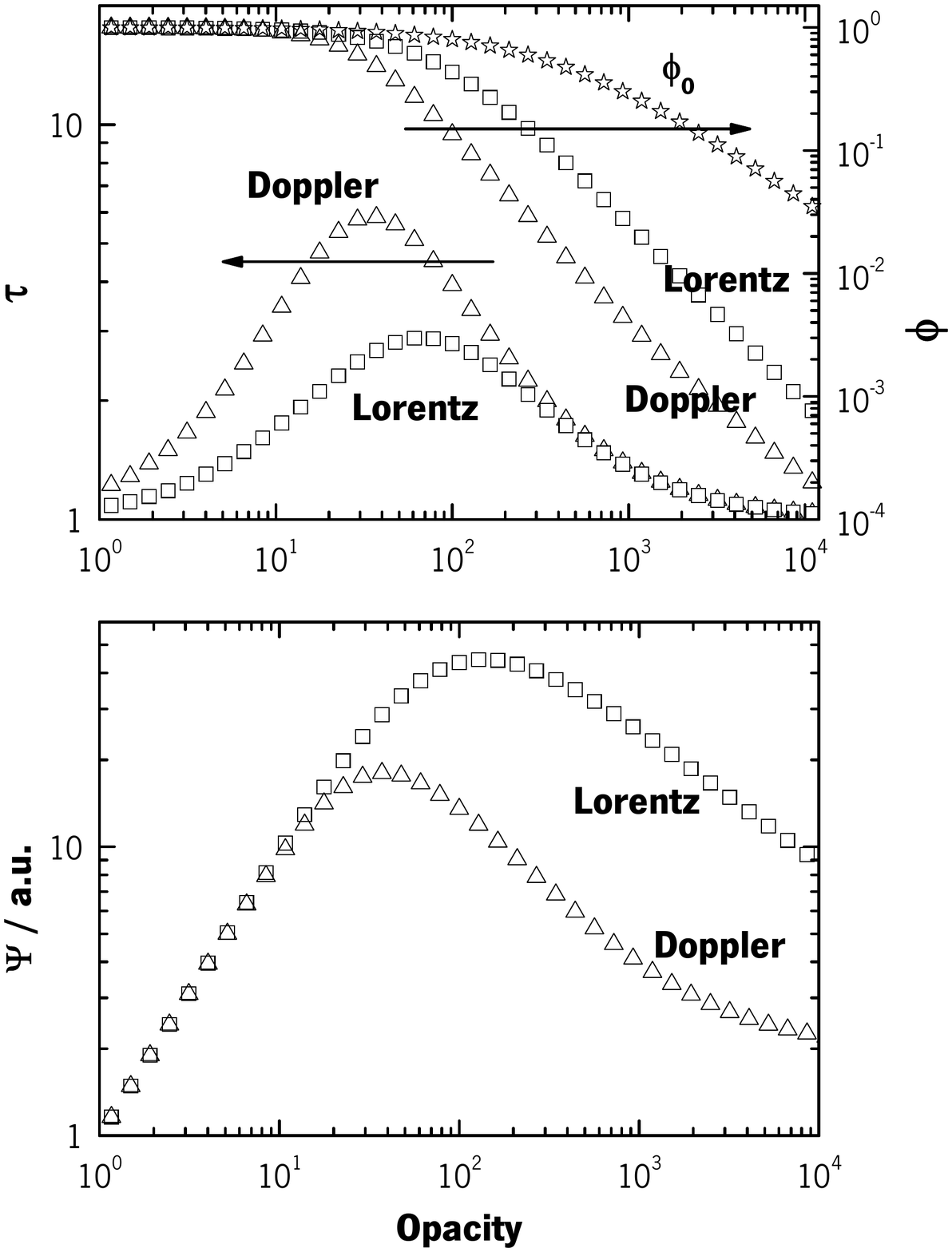}
  \end{center}
  \caption{\label{Fig-Lamp}
    Ensemble relaxation parameters as a function of overall opacity for Doppler and
    Lorentz lineshapes. The upper part shows the Reemission yield~$\phi$ and mean
    scaled lifetime~$\tau$ while the lower part shows the predicted relative efficiency for lighting
    applications estimated from~\eref{phi_lamp}. $\phi_{0}$ is implicitly dependent
    upon opacity~(upper part and text).}
\end{figure}

The actual behaviour of a lamp or a PDP can be quite involved since an increase in opacity means an
increase of partial vapour pressure~(the external dimensions of the device are fixed) and this
induces several changes whose influence on the overall performance can be contradictory. The higher
opacity means higher light throughput \textit{as long as} the increase in the trapping efficiency
does not substantially increase thermal degradation. The higher the opacity the more efficient the
trapping~(higher recapture probabilities) but, on the other hand, the resulting higher collisional
effective rate constants could give rise to a smaller intrinsic~$\phi_{0}$ and render the Lorentz
distribution only approximately valid. A reduced~$\phi_{0}$ value and a shift of the Voigt
distribution towards a more Lorentz-like profile tend by themselves to decrease trappping
efficiency and increase light throughput. To have a simple idea of the effect, and due to the
paramount importance of the reemission quantum yield~$\phi_{0}$, a series of results were made for
both limiting Doppler and Lorentz distributions with the radiative quantum yield given
by~$\phi_{0}=\frac{\Gamma _{r}}{\Gamma_{r}+\Gamma_{q}}$, where the quenching rate constant was
assumed in a first order approximation to be linear with the cell opacity~($\Gamma_{q} \equiv k \,
r$, with the numerical values $\Gamma_{r}=10^{7}$~s$^{-1}$ and $k=2.5\times 10^{4}$~s$^{-1}$). This
corresponds to the well known Stern-Volmer equation for dynamical quenching by binary collisions
for unitary intrinsic radiative yield in the absence of collisions. The results are shown
in~\fref{Fig-Lamp} which is judged more representative of the actual lamp behaviour than the
results in figures~\ref{Fig-Dop} and \ref{Fig-Lor}. \Fref{Fig-Lamp} shows that, ultimately, a
delicate balance will dictate the best operation conditions which manifest themselves in the peaks
of the~$\Psi$ values. Of course, the results are very approximate since the assumed functional
dependence of~$\phi_{0}$ is only approximate and, even within the CFR two-level Lorentzian
assumption, one should use an opacity dependent characteristic width of the Voigt distribution.
Nevertheless,~\fref{Fig-Lamp} allows the discussion of the qualitative behaviour emphasizing
physical insight without the additional burden of fine grained details. It shows how critical the
spectral distribution shape and the quantum reemission yield are. For~$\phi_{0}$ values
sufficiently close to one, an increase in opacity corresponds to an increase in lighting efficiency
due to the increased initial excitation number density. Due to the strong dependence of trapping on
the~$\phi_{0}$ value, as soon as this value starts to be significantly smaller than one, trapping
means an much increased importance of thermal degradation~(compare the difference between the
quantum and the ensemble yield or the reduced overall relaxation lifetime in the upper part
of~\fref{Fig-Lamp}). From some point onwards this will be more important than the increase in
initial excitation due to a higher vapour pressure, which originates optimal operation conditions,
giving the best possible lighting efficiency. \Fref{Fig-Lamp} also shows that the Doppler
distribution is associated with smaller throughput in lighting applications when compared with the
Lorentz case due to the step reduction of the ensemble reemission yield with the increase of the
overall opacity. This is of course related with the use of an inert gas filling to render
collisions more important~(increasing the Lorentz character of the spectral distribution and
reducing trapping efficiency) in fluorescence lamps.

\begin{figure}
  \begin{center}
  \includegraphics[width=12.5cm,keepaspectratio=true]{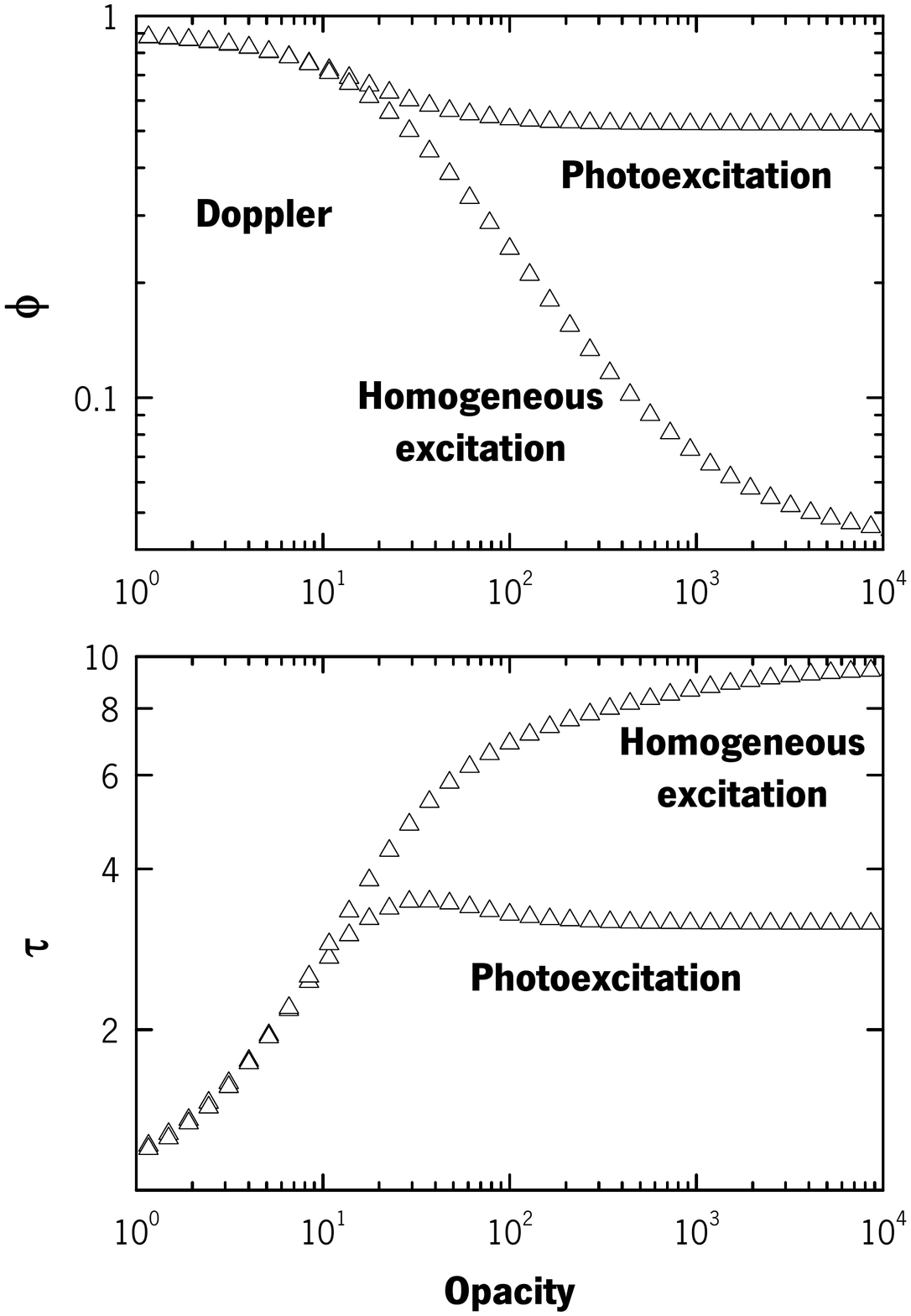}
  \end{center}
  \caption{\label{Fig-HomoPhoto}
    Reemission yield~$\phi$ and mean scaled lifetime~$\tau$ for the Doppler lineshape with $\phi_0=0.90$,
    for homogeneous and external photoexcitation~(with the reabsorption
    undistorted line).}
\end{figure}

Finally,~\fref{Fig-HomoPhoto} shows the predicted ensemble relaxation parameters for both
homogeneous and photoexcitation as a function of overall opacity. Up to opacities of the order
of~$10$ no significant difference exists~(photoexcitation is able to penetrate well deep into
sample cell). But, for higher opacities, the importance of trapping continues to increase
indefinitely for homogeneus excitation while it levels off for photoexcitation, a point to be
revisited in~\sref{Spatial distribution} when discussing spatial distribution functions.

\subsection{Steady-state spectra}\label{Steady-state spectra}

\begin{figure}
  \begin{center}
  \includegraphics[width=12.5cm,keepaspectratio=true]{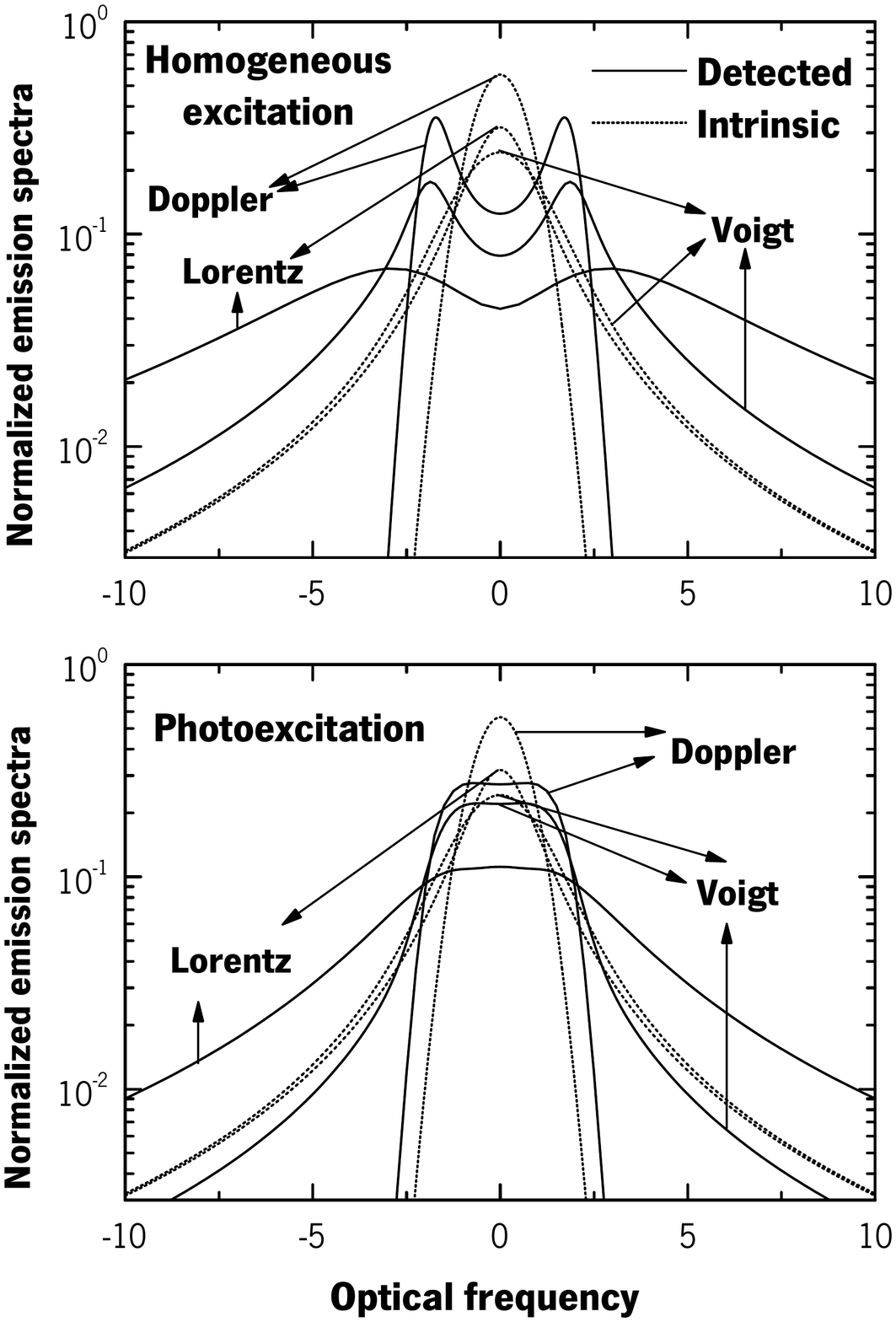}
  \end{center}
  \caption{\label{Fig-SpectraSS}
    Normalized steady-state spectra for Doppler, Lorentz, and~$a=0.1$ Voigt
    lineshapes for an overall opacity of~$100$ and~$\phi_0=0.90$. The photoexcitation
    case corresponds to both excitation and detection from the left cell wall
    using for photoexcitation the reabsorption undistorted line.}
\end{figure}

\Fref{Fig-SpectraSS} shows the estimated normalized spectral distribution in steady-state
conditions for Doppler, Lorentz and Voigt lineshapes for both primary homogeneous and
photoexcitation. The motivation for the homogeneous case is the excitation along the axis of a
fluorescence lamp for lighting applications. It shows the well known self-reversal of spectral
lines due to the higher attenuation of core optical frequencies. For photoexcitation there is a
balance between reduced penetration of external excitation and higher attenuation at core
frequencies which dictates a flattening of the spectra near the line center~(of course, for left
wall photoexcitation and right wall detection there is a self-reversal higher than the one for
homogeneous excitation; not shown). In both cases, there is a considerable broadening of the
detected spectra and the Voigt distribution has an intermediate character between core Doppler-like
and wings Lorentz-like.

\subsection{Spatial distribution}\label{Spatial distribution}

\begin{figure}
  \begin{center}
  \includegraphics[width=12.5cm,keepaspectratio=true]{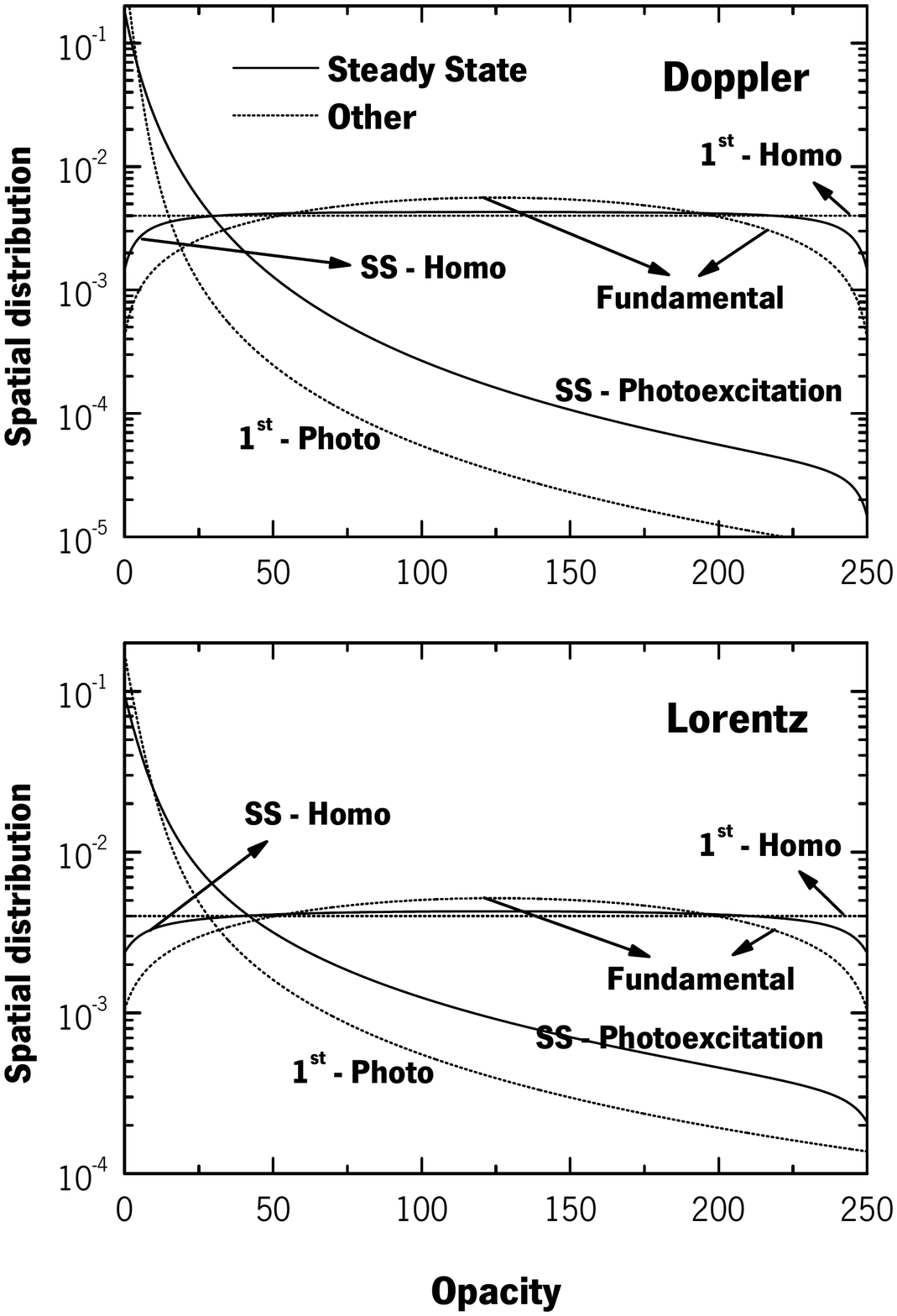}
  \end{center}
  \caption{\label{Fig-Dst}
    Normalized spatial distributions of excitation in steady-state~(SS) conditions for
    Doppler and Lorentz lineshapes for an overall opacity of~$250$ and~$\phi_0=0.90$. It is
    also shown the primary excitation~(homogeneous or photoexcitation) as well as the
    fundamental mode distributions. The photoexcitation case corresponds to both excitation
    and detection from the left cell wall using for photoexcitation the
    reabsorption undistorted line.}
\end{figure}

\Fref{Fig-Dst} shows the spatial distributions for both limiting cases of Doppler and Lorentz
distributions. This figure shows that the fundamental mode spatial distribution~(limiting case for
a relaxed, non-changing spatial distribution and thus independent upon the original excitation)
could only give a reasonable approximation of the steady-state distribution for the homogeneous
excitation case; for photoexcitation it is more convenient to choose the spatial distribution of
the first generation species as a first approximation to the overall distribution. This illustrates
the well known procedure of, whenever approximating the actual trapping dependent behaviour by the
monoexponenial fundamental mode~(easier to obtain by a variational procedure or given by Holstein's
asymptotic approximations), design the experimental setup to mimic as much as possible the
fundamental mode spatial distribution~(symmetrical and well spread into the bulk of sample cell)
with the external excitation. This can be accomplished with photoexcitation of high opacity samples
using strongly detuned external radiation.

\Fref{Fig-Dst} shows some of the difficulties of quantifying trapping simply by using the
fundamental mode, a point further illustrated in~\tref{Table-Fundamental}. Several conclusions can
be draw from its data: (i)~Doppler distributions render trapping much more efficient and thus its
fundamental mode contribution is always much higher than for the Lorentz case, (ii)~the spatial
spreading for Doppler is smaller, giving rise to higher generation number for the fundamental mode,
(iii)~the use of the fundamental mode alone for Lorentz distributions~(and therefore, albeit with a
lesser degree, for Voigt) is never justified and (iv)~to approximate the actual behaviour for
photoexcitation to the fundamental mode is never justifiable.

Two additional points related with the common practice of using only the fundamental mode to take
into account trapping distortions should be stressed out. First, the fundamental mode is the
slowest decaying possible and is located well (and symmetrical) into sample cell. To substitute the
whole of the ensemble dynamics for the fundamental mode alone will always overestimate the lifetime
and underestimate the reemission yield~(spatial distribution giving the highest possible trapping
efficiency) thus introducing a systematic error. Secondly, the use of the fundamental mode alone is
too often misunderstood with the use of the asymptotic approximations proposed by
Holstein~\cite{Holstein1947:PhysRev.72.1212}, only valid in the high opacity limit and for ideal
geometries~\cite{Molisch1998:book}. The MSR has a clear cut advantage in this respect since it
allows an easy estimation of the fundamental mode as the one corresponding to a nonchanging spatial
distribution, irrespective of opacity and geometry.

\begin{table}
  \caption{\label{Table-Fundamental}
  Fundamental mode contribution to the reemission yield~$\phi$ and mean scaled lifetime~$\tau$.
  Also shown is the approximate generation number corresponding to the fundamental mode~($m$),
  for a $10^{-6}$ fractional tolerance to consider a non-changing spatial distribution~(see text).
  In all cases, $\phi_{0}=1$.}

  \begin{tabular}{@{}*{13}{c}}

  \br

  & \centre{6}{Homogeneous} & \centre{6}{Photoexcitation} \\ \ns \ns
  & \crule{6} & \crule{6} \\
  Opacity & \centre{3}{Doppler} & \centre{3}{Lorentz} & \centre{3}{Doppler} & \centre{3}{Lorentz}
  \\ \ns \ns
  & \crule{3} & \crule{3} & \crule{3} & \crule{3} \\ \ns \ns
  & $m$ & $\phi$ & $\tau$ & $m$ & $\phi$ & $\tau$ & $m$ & $\phi$ & $\tau$ & $m$ & $\phi$ & $\tau$ \\

  \br

  $10$ & $10$ & $6$\% & $20$\% & $7$ & $1$\% & $3$\% & $10$ & $3$\% & $14$\% & $7$ & $1$\% & $3$\% \\

  $100$ & $70$ & $10$\% & $40$\% & $30$ & $0.5$\% & $3$\% & $100$ & $2$\% & $20$\% & $35$ & $0.1$\% &
  $1$\% \\

  $1000$ & $200$ & $25$\% & $60$\% & $80$ & $1$\% & $6$\% & $550$ & $0.2$\% & $6$\% & $100$ & $0.1$\% &
  $1$\% \\

  \br

  \end{tabular}

\end{table}

The spatial distribution functions presented in this section draw some further insight into the
previous results of~\fref{Fig-HomoPhoto}. The leveling of the lifetime and reemission yield at
higher opacities for the photoabsorption case corresponds to absorption of external excitation
complete within a layer smaller than the overall opacity making the ensemble relaxation effectively
insensitive to the overall opacity. This approaches well the conditions of semi-infinite geometry,
under the time scale of ensemble complete relaxation.

\section{Conclusions}\label{Conclusions}

This instalment presented a unidimensional computational model study for complete frequency
redistribution linear incoherent atomic radiation trapping illustrating the numerical
implementation of the stochastic model developed previously. It illustrates the advantages of the
multiple scattering representation~(MSR) over the Holstein expansion, based on physical insight and
computation feasibility at an elementary level. Holstein's ansatz has significant shortcomings when
compared to the equivalent alternative of~MSR: (i)~Holstein spatial modes are unphysical except for
the fundamental, (ii)~their estimation is computationally much more troublesome than the simple
algorithms used in this work, (iii)~the wide spread use of original Holstein expressions for the
fundamental mode are only valid in the asymptotic limit of high opacities while~MSR allow an easy
estimation of the fundamental at any opacity value and (iv)~the higher Holstein modes are difficult
to obtain while for Lorentz-like spectral distributions we found that their contribution must be
always taken into account~(the fundamental mode contribution to ensemble relaxation being always
small; higher Holstein modes correspond in the~MSR language to small number generations and are
easy to obtain with the stochastic formulation presented).

The dependence of the ensemble reemission yield and lifetime, relative efficiency for lighting
applications, steady-state spectral and spatial distributions on quantum yield, opacity and
homogeneous or external photoexcitation are discussed at length for the Doppler, Voigt and Lorentz
lineshapes. We quantify the contribution of the nonchanging fundamental mode and found troublesome
using uniquely this mode for Voigt and Lorentz like spectra. The results should appeal to a broad
audience and provide insight in a wide range of more realistic situations in an atomic as well as
in an astrophysical context.

Several possible developments of the general framework presented in this work are shown in
Appendices~B to~D. In the first, we give an outlook of the implementation details that we have
found critical for a Monte Carlo simulation alternative of the Markov algorithm. The Monte Carlo
and Markov approaches constitute in fact two general purpose algorithms for incoherent radiation
propagation problems. Each has in own advantages and shortcomings. The Monte Carlo constitute a
simulation of particle like trajectories while the Markov model quantifies the evolution of
\textit{mean} probabilities. We advocate the second alternative in all but the more demanding
cases~(detailed 3D geometries, realistic multi-level atomic models, partial frequency
redistribution and polarization dependent radiation transport, issues not addressed here). In
~\ref{Unidimensional models for radiation transport}, we give a brief description of a more
realistic but nevertheless still unidimensional Markov implementation of a radiation transport
model, especially appealing for plane-parallel stellar atmosphere theories. Finally, in
~\ref{Specularly reflecting walls} we give an analytical modification of the Markov algorithm for
multiple specular reflections in boundaries. The appendices could be useful for more advanced
projects.


\ack

This work was supported by Funda\c{c}\~{a}o para a Ci\^{e}ncia e Tecnologia~(Portugal) and
Universidade do Minho~(Portugal) within project~REEQ/433/EEI/2005. It also used computational
facilities bought under project~POCTI/CTM/41574/2001, funded by FCT and the European Community
Fund~FEDER. A.R.~Alves-Pereira acknowledges~FCT funding under the reference~SFRH/BD/4727/2001.
E.~Nunes-Pereira acknowledges the critical reading of the manuscript by M.~Besley~(Centro de
F\'{i}sica, Universidade do Minho).


\appendix

\section{Numerical Voigt distribution}\label{Numerical Voigt distribution}
\setcounter{section}{1}

The numerical evaluation of the Voigt spectral distribution can be troublesome, as one can easily
judge from the large number of approximations that have been published in the last decades
balancing precision and computational speed~(see~\cite{Molisch1998:book} and references therein).
This is especially true for the wings of the distribution in case of trapping since the photons can
most easily escape via the wings, especially in high opacity vapours. However, given the current
desktop computing capabilities, the numerical (careful) direct integration of the defining equation
is perfectly adequate and the use of approximations to reduce computation time is not justifiable
any more. The difficulty in the direct numerical integration cames from the behaviour of the
integrand function: it differs from zero over two width scales, a broad scale centered in
zero~(corresponding to the exponential term in the numerator of the integrand) and a much narrower
one centered in a frequency corresponding to the Voigt frequency~(associated with the difference
term in the denominator). The integration domain should thus be broken into smaller domains and an
automatic adaptative integration algorithm should be used in each subdomain always starting at the
integrand function maximum and with a initial stepsize adapted to the local scale of variation of
the integrand~\cite{NR-F77-2nd:book&NR-F90-2nd:book}. We have used the~$400$ central frequencies
for the central broad feature and a~$0.4$ frequency width for the floating narrow peak.
Integrations further away from the central core were analytically mapped from an infinite to a
finite integration range. In order to decrease the time for repetitive Voigt functions evaluations,
the Voigt distribution was previously computed in a given table of frequency values and, whenever
necessary, cubic spline interpolated~\cite{NR-F77-2nd:book&NR-F90-2nd:book}. We used a linear scale
in the core~(frequency range up to $100$ with a~$5\times 10^{-2}$ spacing) and a log scale in the
wings~(frequency range from~$100$ to $10^{8}$ with~$1\times 10^{-3}$ $log_{10}$ spacing). Natural
cubic spline was not necessary since the derivatives of the Voigt distribution in the end points
are analytical.

\section{Monte Carlo simulation}\label{Monte Carlo simulation}

The mean reabsorption and escape probabilities can be estimated either with a Markov chain
algorithm or with Monte Carlo simulation mimic of the experiment. We will outline the computational
details we found critical for the Monte Carlo alternative and give our best advice on each method
strengths and limitations and under which physical conditions the Monte Carlo is especially
adequate.

The Monte Carlo~(MC) method makes a direct simulation of the trapping process using particle-like
trajectories for radiation in cell~\cite{MC-CRC-1990:book}. The initial excitation coordinate is
randomly chosen corresponding to either homogeneous or external photoexcitation. The (re)emission
coordinate is the same as the absorption one and, after emission, a random direction and optical
frequency~$x$ must be chosen from the appropriate distributions. The photon path in cell is
followed and a reabsorption coordinate drawn from the Beer-Lambert exponential distribution with
absorption coefficient given by~$\Phi \left( x \right)$. This should be then tested for escape from
cell; if the photon escaped, another excitation trajectory should be initiated from the first
generation, otherwise the simulation must continue by increasing generation number and repeating.
Appropriate counters keep track of the actual number of trajectories giving rise to $n^{th}$
generation species~(the ratio to the total number of trajectories initiated for first generation
gives the population efficiencies), and the mean escape probabilities~(for each generation number,
excitation coordinate~$r$ and optical frequency~$x$, the escape probabilities counters are
incremented with $\me ^{-\Phi \left( x\right) r}$ for left escape; compare~\eref{EscapeOmega1D}).
To decrease the computation burden, an importance sampling method should be used always assuming a
unit intrinsic reemission yield and after the influence of the actual value of the yield introduced
analytically, as described in the main text. The spatial distribution functions for each generation
are most easily obtained for a discretized cell just by keeping track of the number of excited
species in each cell bin but it is important to acknowledge the fact that this binning is used only
for graphical representation purposes since the mean escape probabilities are computed from the
actual spatial coordinates. This is an an important difference relative to the Markov algorithm
since this last case effectively makes a simulation on a lattice model from the very onset.

The complete Monte Carlo description lacks only the fine details of the implementation of the
transformation method to obtain non-uniform deviates~\cite{NR-F77-2nd:book&NR-F90-2nd:book}. These
are used in the simulation to obtain excitation coordinates and optical frequencies. The excitation
coordinates are either homogeneous for initial excitation~(trivial) or drawn form the Beer-Lambert
law, giving the reabsorption coordinate relative to emission point in an opacity scale for a given
absorption coefficient at each optical frequency~$\Phi\left(x\right)$. Since Beer law corresponds
to an exponential distribution, the random deviates are analytically given
by~$-ln(y)/\Phi\left(x\right)$, where~$y$ is a uniform deviate. For the optical frequencies~$x$, we
must distinguish between Doppler, Lorentz and the Voigt lineshapes. The Doppler and Lorentz
distributions have analytical inverses and are therefore simpler. Lorentz deviates are given
by~$x=\tan \left[ \pi \left( y-1/2\right) \right] $ while Doppler correspond to zero mean~$1/2$
variance normal deviates. The reference~\cite{NR-F77-2nd:book&NR-F90-2nd:book} only gives directly
a routine for zero mean unit variance deviates but it is easy to derive a general purpose formula
for zero mean, arbitrary~$\sigma $ variance along the same lines of the used Box-Mueller algorithm.
The Voigt case is more troublesome since the transformation method must be implemented numerically.
The most straightforward implementation of this is to use a cubic spline of the cumulative
distribution function, truncated to a very high frequency (se have obtained good results with an
upper limit of $10^{8}$, on a logarithmic scale in the wings).

For the simple model study of this work, the Markov algorithm has distinctive advantages over the
Monte Carlo alternative since it considers directly the evolution of \textit{mean} excitation
probabilities. These can be obtained with a fast algorithm and this is very important for trapping
due to the critical need to proper quantify the fundamental mode. The estimation of the fundamental
mode in MC is much more difficult due to the characteristic slow convergence of MC estimates and to
the fact that MC simulations must be made with a maximum generation number specified at start. This
is especially important for high opacity cases since in this case the MC computation time can be
prohibitly large and we have found that a small error in the fundamental mode can be greatly
amplified in overall relaxation parameters. MC is nevertheless more versatile with respect to
generalizations to either more realistic geometries and atomic models as well as to include
polarization effects. The influence of geometry deserves a small note. For realistic tridimensional
geometries the demands introduced by a sufficiently fine spatial discretization in the Markov
procedure can compromise the practical feasibility with current desktop technology. For~MC there is
no significant added overhead computational complexity since the level of discretization only
changes the visual resolution of the spatial distribution functions. Balancing the shortcomings of
both approaches, our advice would be to use Markov whenever possible. For 3D geometries, a possible
procedure would be:~(i)~to identify the smallest dimension in cell, (ii)~to do a 1D Markov
estimation of the fundamental mode generation number for that smallest opacity and (iii)~finalize
by MC simulating the actual geometry until that generation number~(assumed to describe well the
fundamental mode).

\section{Unidimensional models for radiation transport}\label{Unidimensional models for radiation transport}

The unidimensional implementation of radiation migration used in this work is certainly very naive.
An alternative~1D possibility nevertheless exists that does not imply a qualitative increase in
computation complexity and which is more realistic. It introduces some approximations, albeit of a
different nature. To keep the unidimensional formulation of radiation migration, one can represent
trapping as a function of a characteristic distance measured in the perpendicular to some known
surface. This corresponds to the well known cases of plane parallel stratified stellar
atmospheres~\cite{Mihalas1978:book} or the idealized one-dimensional geometries in laboratory scale
atomic vapours~\cite{Molisch1998:book}. In the Markov algorithm, the one-step transition
probabilities should be modified in order to take into account all the possibilities of transition
between any two coordinates with the same difference in perpendicular distances to the reference
surface. This can be done for homogeneous three-dimensional space, by projecting the transition
probability on an arbitrary axis~(the one used for the~1D geometry). Instead
of~\eref{Transition-pij}, one obtains~\cite{SuperdiffusionI:JCP.125.174308}

\begin{equation}
  p^{ij} \simeq \frac{1}{2}h\int_{-\infty }^{+\infty } \Phi^{2}\left( x\right)
  {\rm E}_{1}\left( \left\vert \bi{r}_{i}-\bi{r}_{j}\right\vert \, \Phi \left( x\right) \right)
  \dif x
  \mbox{,}
\end{equation}

where~${\rm E}_{1} \left( x\right) $ is the exponential integral function, defined as ${\rm E}_{1}
\left( x\right) = \int_{x}^{+\infty } \frac{\me ^{-u}}{u} \, \dif u$. No further modifications of
the Markov stochastic algorithm are needed.

\section{Specularly reflecting walls}\label{Specularly reflecting walls}

Another case of practical interest is a vapour cell with partially reflecting
walls~\cite{Molisch1998:book}. For the unidimensional geometry and for specularly reflecting walls,
the Markov algorithm is amenable to a simple modification since the absorption, transition and
escape probabilities can be cast as series which have analytical
representations~\cite{Nunes-Pereira1995:JLumin.63.259}.

Consider a unidimensional geometry with zero at the left and~$r_{max}$ maximum overall opacity in
the right wall. The corresponding reflectances are assumed constant and will be represented
as~$R_{L}$ and $R_{R}$. The stochastic interpretation of the Beer-Lambert law gives for the
absorption and escape mean probabilities, after an~$r$ optical pathlength, $\Phi \left( x\right) \,
\me ^{-\Phi \left( x\right) r}$ and $\me ^{-\Phi \left( x\right) r}$, respectively. $\Phi \left(
x\right)$ is just the mean probability of reemission of a photon with frequency~$x$. With all of
this in mind, the Markov transition and escape probabilities can be reformulated taking into
account the possibility of multiple specular reflections of the geometry boundaries. For the mean
transition between Markov states~$i$ and $j$, instead of~\eref{Transition-pij} one can write

\begin{equation}
\label{Transition-pij-R-1} p^{ij} \simeq \frac{1}{2}h\int_{-\infty }^{+\infty }
f^{ij}\left(x,R_{L},R_{R}\right) \dif x
  \mbox{,}
\end{equation}

with the integrand function given by

\begin{eqnarray}
\label{Transition-pij-R-2} f^{ij}\left(x,R_{L},R_{R}\right) = \Phi ^{2}\left( x\right) \me ^{-\Phi
\left( x\right) \left\vert r_{i}-r_{j}\right\vert } + \\
\quad + \Phi ^{2}\left( x\right) \me ^{-\Phi \left( x\right) r_{i}} \, \me ^{-\Phi \left( x\right)
r_{j}} \, R_{L} \left[ \sum_{n=0}^{+\infty } R_{L}^{n}R_{R}^{n} \left( \me ^{-\Phi \left( x\right)
r_{\max }}\right) ^{2n}\right] + \nonumber \\
\quad + \Phi ^{2}\left( x\right) \me ^{-\Phi \left( x\right) r_{i}} \, \me ^{-\Phi \left( x\right)
\left( r_{max}-r_{j} \right) } \, R_{L} R_{R} \left[ \sum_{n=0}^{+\infty } R_{L}^{n}R_{R}^{n}
\left( \me ^{-\Phi \left( x\right)
r_{\max }}\right) ^{2n+1}\right] + \nonumber \\
\quad + \Phi ^{2}\left( x\right) \me ^{-\Phi \left( x\right) \left( r_{max}-r_{i} \right) } \, \me
^{-\Phi \left( x\right) \left( r_{max}-r_{j} \right)} \, R_{R} \left[ \sum_{n=0}^{+\infty }
R_{L}^{n}R_{R}^{n} \left( \me ^{-\Phi \left( x\right)
r_{\max }}\right) ^{2n}\right] + \nonumber \\
\quad + \Phi ^{2}\left( x\right) \me ^{-\Phi \left( x\right) \left( r_{max}-r_{i} \right) } \, \me
^{-\Phi \left( x\right) r_{j}} \, R_{L} R_{R} \left[ \sum_{n=0}^{+\infty } R_{L}^{n}R_{R}^{n}
\left( \me ^{-\Phi \left( x\right) r_{\max }}\right) ^{2n+1}\right] \nonumber
  \mbox{.}
\end{eqnarray}

The first line gives the transition probability due to direct absorption prior to any reflection.
The other lines give the additional absorption probability in state~$j$ after at least one
reflection in the cell walls. The second and third lines correspond to emission from state~$i$ to
the \textit{left} while the fourth and fifth correspond to emission to the \textit{right}. Finally,
the second and fifth lines consider photon absorption in state~$j$ with the radiation coming
\textit{from the left} of this state while the third and fourth lines correspond to absorption
\textit{from the right}.

Rearranging, one eventually obtains

\begin{eqnarray}
\label{Transition-pij-R-3} f^{ij}\left(x,R_{L},R_{R}\right) = \Phi ^{2}\left( x\right) \me ^{-\Phi
\left( x\right) \left\vert r_{i}-r_{j}\right\vert } + \\
\quad + \Phi ^{2}\left( x\right) \frac{\me ^{\Phi \left( x\right) r_{max}}}{\me ^{2 \Phi \left(
x\right) r_{max}}-R_{L}R_{R}} \left\{ R_{L} \me ^{-\Phi \left( x\right) \left( r_{max}+r_{i}+r_{j}
\right) } + \right. \nonumber \\
  \quad \quad \left. + R_{R} \me ^{-\Phi \left( x\right) \left(r_{max}-r_{i}-r_{j} \right) } +
  R_{L}R_{R} \left( \me ^{-\Phi \left( x\right) \left( r_{max}+r_{i}-r_{j} \right) } + \me ^{-\Phi
\left( x\right) \left( r_{j}-r_{i} \right) } \right) \right\} \nonumber
  \mbox{.}
\end{eqnarray}

Using the same procedure, the left escape probability is given by

\begin{eqnarray}
\label{Transition-q-R}   q^{i\Omega }\left( x\right) \simeq \frac{1}{2}
  \me^{-\Phi \left( x\right) r_{i} } \left( 1-R_{L} \right) \left[ \sum_{n=0}^{+\infty }
  R_{L}^{n}R_{R}^{n} \left( \me ^{-\Phi \left( x\right) r_{\max }}\right) ^{2n}\right] + \\
  \quad + \frac{1}{2}
  \me^{-\Phi \left( x\right) \left( r_{max}-r_{i} \right) } \left( 1-R_{L} \right) R_{R}
  \left[ \sum_{n=0}^{+\infty }
  R_{L}^{n}R_{R}^{n} \left( \me ^{-\Phi \left( x\right) r_{\max }}\right) ^{2n+1}\right] \nonumber \\
  \simeq \quad \frac{1}{2} \left( 1-R_{L} \right)
  \frac{\me ^{\Phi \left( x\right) r_{max}}}{\me ^{2 \Phi \left( x\right) r_{max}}-R_{L}R_{R}}
  \left\{ \me ^{-\Phi \left( x\right) \left( r_{i}-r_{max} \right) } +
  R_{R} \me ^{-\Phi \left( x\right) \left( r_{max}-r_{i} \right) } \right\} \nonumber
  \mbox{.}
\end{eqnarray}

The first line is the \textit{left} escape, given that the emission was is state~$i$ to the
\textit{left} while the second is the \textit{left} escape for initial \textit{right} emission.

Finally, multiple reflections can also change the initial spatial distribution in the case of
external photoexcitation. For the case of photoexcitation from the \textit{left} side with the
undistorted resonance line, \eref{Initial-p1i} should be modified into

\begin{equation}
  \label{Initial-p1i-R-1}
  p_{1}^{i} \simeq h\int_{-\infty }^{+\infty } f_{1}^{i}\left(x,R_{L},R_{R}\right) \dif x
  \mbox{,}
\end{equation}

with $f_{1}^{i}\left(x,R_{L},R_{R}\right)$ given by

\begin{eqnarray}
  \label{Initial-p1i-R-2}
  f_{1}^{i}\left(x,R_{L},R_{R}\right) =
  \Phi ^{2}\left( x\right) \me^{-\Phi \left( x\right) r_{i} } \left[ \sum_{n=0}^{+\infty }
  R_{L}^{n}R_{R}^{n} \left( \me ^{-\Phi \left( x\right) r_{\max }}\right) ^{2n}\right] + \\
  \quad + \Phi ^{2}\left( x\right) \me^{-\Phi \left( x\right) \left( r_{max}-r_{i} \right) } R_{R}
  \left[ \sum_{n=0}^{+\infty }
  R_{L}^{n}R_{R}^{n} \left( \me ^{-\Phi \left( x\right) r_{\max }}\right) ^{2n+1}\right] \nonumber \\
  \quad = \Phi ^{2}\left( x\right)
  \frac{\me ^{\Phi \left( x\right) r_{max}}}{\me ^{2 \Phi \left( x\right) r_{max}}-R_{L}R_{R}}
  \left\{ \me ^{-\Phi \left( x\right) \left( r_{i}-r_{max} \right) } +
  R_{R} \me ^{-\Phi \left( x\right) \left( r_{max}-r_{i} \right) } \right\} \nonumber
  \mbox{.}
\end{eqnarray}

The first line is the absorption in state~$i$ of radiation propagating from \textit{left} while the
second is the absorption of radiation coming from the \textit{right}.

For specular reflection in this unidimensional model the advantages of the Markov algorithm over
the alternative Monte Carlo are even more important than for the case in which reflection in
boundaries is not considered. The multiple reflections are taken into account analytically for the
Markov case while in the Monte Carlo simulation the increase in computation time with the
increasing importance of reflections can be very high. Nevertheless, the increase in trapping
efficiency with the added possibility of specular reflection is only significant if both reflection
coefficients are important. In fact, a 1D geometry in which one of the boundaries is perfectly
reflective while the other has a zero reflection coefficient is equivalent to a 1D sample of twice
the size of the original one.


\vspace{1cm}





\end{document}